# An Integrated Platform for Studying Learning with Intelligent Tutoring Systems: CTAT+TutorShop

Vincent Aleven[1], Conrad Borchers[1], Yun Huang[2], Tomohiro Nagashima[3], Bruce McLaren[1], Paulo Carvalho[1], Octav Popescu[1], Jonathan Sewall[1] and Kenneth Koedinger[1]

[1]Carnegie Mellon University, {aleven,cborcher,bmclaren,pcarvalh,octav,sewall,krk}@cs.cmu.edu

[2]Austral University of Chile, yun.huang@uach.cl

[1]Saarland University, nagashima@cs.uni-saarland.de

**Abstract** Intelligent tutoring systems (ITSs) are effective in helping students learn; further research could make them even more effective. Particularly desirable is research into how students learn with these systems, how these systems best support student learning, and what learning sciences principles are key in ITSs. CTAT+Tutorshop provides a full stack integrated platform that facilitates a complete research lifecycle with ITSs, which includes using ITS data to discover learner challenges, to identify opportunities for system improvements, and to conduct experimental studies. The platform includes authoring tools to support and accelerate development of ITS, which provide automatic data logging in a format compatible with DataShop, an independent site that supports the analysis of ed tech log data to study student learnings. Among the many technology platforms that exist to support learning sciences research, CTAT+Tutorshop may be the only one that offers researchers the possibility to author elements of ITSs, or whole ITSs, as part of designing studies. This platform has been used to develop and conduct an estimated 147 research studies which have run in a wide variety of laboratory and real-world educational settings, including K-12 and higher education, and have addressed a wide range of research questions. This paper presents five case studies of research conducted on the CTAT+Tutorshop platform, and summarizes what has been accomplished and what is possible for future researchers. We reflect on the distinctive elements of this platform that have made it so effective in facilitating a wide range of ITS research.

## 1 Introduction

Intelligent tutoring systems (ITSs) [42] are a form of ed tech proven to be highly effective in helping students learn [10, 31, 38]. Although there is a large research literature on learning with ITSs, further research is highly desirable. The better we understand how students learn with this technology, what features make it effective, and how to carry out data-driven improvement cycles, the better we will be able to enhance the engineering of these systems and make them even more effective. Further, developing a deeper understanding of student learning in highly-scaffolded environments (as ITSs are) is interesting in its own right from a learning sciences perspective. What makes ITSs particularly attractive vehicles for learning sciences research is that they can be used in a full research lifecycle. That lifecycle includes mining log data from students' interactions to discover learner challenges (e.g., through methods from Educational Data Mining) and identify opportunities for system improvements, creating alternative designs aimed at bringing about improvements, and testing them in A/B experiments. ITSs can serve to address to a wide range of research questions, not only questions about how best to implement particular ITS functionality, but also about what learning sciences principles are key in the context of the high scaffolding provided by ITSs

This paper reports on the use of CTAT+Tutorshop as a research platform for investigating learning sciences questions around learning with ITSs, a role it has served for over 20 years. The Cognitive Tutor Authoring Tools (CTAT) were originally developed to facilitate the development of ITSs for both programmers and non-programmers [4]. These tools are largely domain-independent and can be (and have been) applied for start-to-finish tutor development in many task domains and at many levels of instruction, including K-12, colleges, and universities. Tutors built with CTAT have the main features that are characteristic of ITSs: they adaptively guide students in complex learn-by-doing activities with both within-problem guidance (or an "inner loop," or "step loop") [63] and individualized mastery learning [14]. Tutorshop was first conceived as a Learning Management System (LMS) to support the use of CTAT-built tutors in real educational settings such as classrooms and university courses. The latest version of the CTAT authoring tools, which runs on the web, is embedded in Tutorshop, bringing the systems even closer together. This embedding for example facilitates collaborative authoring and publishing tutors on Tutorshop, where they are available for testing and real-world use. Tutors developed on the platform produce rich log data, in DataShop format [28], an independent site that offers many analytics and tools for analyzing student learning. Many major tutoring systems and educational games have been built and used within this platform including: Mathtutor [3], the Genetics Tutor [15], the Stoich Tutor [39], ORCCA [27], Chem Tutor [54], the Signals Tutor [57], Lynnette [34, 60], Gwynnette [44], several incarnations of the Fractions Tutor [16, 49, 51], and Decimal Point [46].

Many of these ITSs were used as part of regular instruction in schools and university courses, but they have also been used in a very large number of research studies (estimated to be 147, see below). Many of these studies tested effects of new tutoring features (e.g., new implementations of learning principles within tutoring systems) or tested the effectiveness of tutoring systems against other forms of instructions, in a range of study designs. Over the lifetime of the CTAT+Tutorshop platform, we have helped researchers become familiar with the platform (through courses, summer workshop, and consulting) and run studies on the platform (through consulting, sometimes by taking a larger role in helping them implement their study). Although past papers have described both CTAT and Tutorshop [4], they have not focused on their use as a research platform. In the current paper, we address that gap.

The combination of CTAT+Tutorshop (together with DataShop) supports research in the first place by facilitating authoring, deploying, and using ITSs in real educational settings. In addition, the tools have some dedicated functionality to facilitate research studies, described in more detail below, including, first and foremost DataShop logging [28]. There are many advantages to having authoring tools integrated in a research platform, even if at first blush, authoring ed tech might seem antithetical to the idea of platform-enabled research – isn't the number one reason for using an existing platform that the researcher does not need to create their own? Arguably, the CTAT+Tutorshop authoring tools facilitate a wide range of research. First, as mentioned earlier, a range of ITSs have been built with these tools that are now available for use in research studies – so researchers do not need to build their own. Interestingly, some ITSs created with CTAT that were primarily created for use in classrooms were also used as a platform for research [6, 15, 27]; the two motivations often go hand in hand. Furthermore, authoring tools facilitate the creation of experimental conditions for use in experimental studies. These conditions can be variations of existing ITSs, authored with the authoring tools. Researchers can also use the authoring tools to create a new ITS to serve as a new platform for research [62].

It is not clear that it matters whether or how one learning-research-enabling platform is different from another, since there is a very large number of research questions to be answered – in that sense, the more high-quality platforms, the merrier. If we nonetheless had to point out the distinguishing characteristics of CTAT+Tutorshop, we would say it is unique in that it focuses on ITSs and integrates (programmer and non-programmer) authoring tools. Close comparisons are Assistments (+ eTrials), OATtutor, and MATHia (+

UPGRADE). The Assistments platform provides authoring tools, but for a restricted class of ITSs (e.g., with one top level question and one way of solving it). OATutor provides mass-production of tutoring content but no authoring tools for non-programmers to customize tutoring interfaces and functionalities [47]. UPGRADE provides some authoring capabilities but does not give researchers access to the full-blown set of MATHia authoring tools, which is proprietary.

The paper is structured as follows: We start out by characterizing ITSs and describing how they are built and deployed on CTAT+Tutorshop. We then describe how the platform, with DataShop, supports research, characterizing the range of research studies that the platform could support. Finally, we describe and illustrate the research that has been done on the platform, including estimates of studies and users, and we present five case studies.

## 2  Background: Intelligent Tutoring Systems with CTAT+Tutorshop

Generally, ITSs support deliberate practice of cognitive skills with feedback, just-in-time instruction (aka hints) and individualized mastery learning [14, 29]. Typically, they support practice of problems that require multiple reasoning steps to solve and that can be solved in multiple ways – that is, have multiple solution paths students may follow. We call the guidance along multi-step solution paths "within-problem guidance," which is meant to be equivalent to VanLehn's [63] notion of an "inner loop" or "step loop." Within-problem guidance enables students to succeed on complex problems, with which they would likely struggle otherwise. As well, with such guidance, students learn better ([64, discussed in [38]). As an additional form of adaptivity, ITSs support personalized problem selection – every learner gets the right (individualized) amount of practice to master the knowledge components (KCs) targeted in the instruction. The adaptive guidance within problems and adaptive problem selection aimed at mastery learning set ITSs apart from other ed tech.

The Cognitive Tutor Authoring Tools (CTAT; [4]) support the start-to-finish development of two types of ITSs, example-tracing tutors, which can be created without programming (e.g., by researchers without a technology background), and model-tracing tutors, which are more suitable for problems with large solution spaces, but require AI programming to create a rule-based cognitive model. Many research studies show that students learn very effectively with CTAT-authored tutors (studies listed in Table 1).

Tutorshop is an LMS dedicated to using CTAT-built tutors in real educational settings (schools, colleges) as well as for personal use. It lets teachers create class lists, assign work to students, and view performance reports for their classes. It allows students to see their assigned work, launch the tutoring system, and view reports about their performance and progress. Tutorshop also allows researchers to create or modify tutors and/or problem sets. It also allows researchers to perform a variety of useful administrative functions related to experiments, such as assigning different activities to students in different conditions, or tracking progress. We note that CTAT-authored tutors can be embedded easily in a range of widely-used LMSs (e.g., OpenEdX, OLI, Canvas, Moodle), where they could be used for research.

A researcher running a study on CTAT+Tutorshop may opt to create a new tutoring system for use in their study or decide to use one of the existing tutoring systems available within Tutorshop. Either way, they will likely create variations of the tutoring system to serve as conditions in the experiment. Once an author has identified the targeted problem types and problems for which to build a tutor and has selected which of the two tutor paradigms to use, key design decisions pertain to:

- **The layout of the tutor interface:** How the problem is divided into subgoals and steps, when subgoals or steps are made visible, whether/how the interface prompts the student for the steps.
- **The knowledge component (KC) model:** How the overall competency to be learned in a given tutor unit is decomposed into KCs, which the tutor uses to track a student's knowledge growth.
- **Adaptive step-level guidance:** The author needs to provide step-level hint messages as well as feedback messages presented upon successful completion of steps or specific errors.

- **Adaptive task selection policy:** An author must select one of the task selection policies available in Tutorshop (such as a fixed problem sequence and individualized mastery learning) or write their own custom task selection policy (through programming).

These design features can be viewed as the "raw materials" from which experimental conditions within ITS can be built – e.g., an author could craft a version of a tutoring system that reflects a learning principle such as "use polite language' and a version that is the same except for the language is "regular" rather than "polite;" or a system with and without worked examples (see below under case studies). Over the years, we added several features specifically added to aid research:

- Data logging built into tutors (this feature has been used in every single research study done with CTAT+Tutorshop)
- Authoring of tests for students, essentially tutors without hints and feedback, for use in pre/post tests
- Ways of configuring the tutor behavior (e.g., several predefined hint, feedback, task selection policies to choose from) which could make for easily-implemented experiments
- Support (i.e., an API) so authors or researchers can craft their own task selection policies, which is useful for studies in which authors test new policies or compare different policies
- Extensible student model, so an author can add (through programming) non-standard student model variables, together with an API for custom "detectors"(e.g., student engagement, disengagement, system misuse, affect, progress, aspects of self-regulated learning, etc. (Holstein et al., 2018); CTAT offers various ways to make tutor behavior contingent on these variables, including custom task selection policies and the embedding of formulas in hints or even in the criteria for correct steps
- Associating a condition with an assignment, so that when experimental conditions are implemented as Tutorshop assignments (as they often are), each student's condition can be logged
- Experimenter accounts, which have all the privileges that researchers typically need

# 3   How CTAT+Tutorshop Can Be Used for Research

The CTAT+Tutorshop platform supports research into how students learn with ITSs, how these systems best support learning, and what learning sciences principles are key in ITSs. The platform supports a research and engineering cycle in which data from an ITS is used to discover regularities and challenges in learning with ITSs, to identify opportunities for system improvement, and to test alternative interventions in experimental studies. Although not all studies go through the full cycle, the cycle is available when the research questions warrant going through it. The paragraphs below elaborate on each cycle stage. Examples of studies in these stages are listed in Table 1 – this table is not exhaustive, as many more studies have been carried out on the CTAT+Tutorshop platform.

**Identify regularities and challenges –** Several studies conducted on the platform focused on Identifying challenges that students experience when learning with a given ITS (or across multiple ITSs) as well as on discovering patterns in students learning with one or more ITSs. These studies involve data mining, using log data captured in DataShop. A common type of analysis is KC model analysis, where the researcher tries to identify a KC model that best captures the distinct skills evident in students' longitudinal performance on problem-solving steps (e.g. [24, 36]).

**Identify opportunities for improvement, redesign –** One study used the results from an analysis of log data to inform a data-driven redesign of the given ITS. This study represents a form of design loop adaptivity, in the terminology of the Adaptivity Grid, a conceptual framework for understanding how learning technologies can adapt to both differences and similarities among learners [2]. Design-loop adaptivity means making a system more adaptive to challenges experienced by *all* learners in a given task domain, based on data from the system. Ideally, the effect of the redesign is evaluated in a close-the-loop study, by comparing it against the original system.

**Intervention studies** – The CTAT+Tutorshop platform has been used for a wide variety of intervention studies, which test effects, typically on students' learning processes and outcomes, of ITS variations, sometimes against business-as-usual control conditions. Students' learning processes are often studied by analyzing tutor log data in DataShop, while learning outcomes are frequently measured with pre- and post-tests, which can be authored with CTAT and administered on the Tutorshop platform. Other intervention studies test **learning with an ITS against external, often non-ITS control conditions (where the latter are typically not** implemented in CTAT+Tutorshop). Examples of such control conditions are a commercially available educational game or doing the same exercises on paper (without tutoring support; [7]). Finally, several studies carried out on the CTAT+Tutorshop platform investigated the **effect of teacher analytics tools such as dashboards or the effectiveness of authoring tools to build ITSs.**

## 4 Research Done with CTAT+Tutorshop

We do not have an exact count of the number of research studies that have been run on the CTAT+Tutorshop platform, nor of the exact number of researchers that have been active on the platform. The reason that obtaining such a count is difficult is that the platform has been used for many different purposes, including tutor development and use of tutors in regular instruction. Also, the platform does not have an explicit representation of a study, and we do not require users to formally describe their purpose. Instead, we give an impression of the amount of research done and the number of active researchers, by counting the datasets in DataShop that come from CTAT tutors and the Tutorshop accounts that created content.

**Table 1:** Research studies on on CTAT+Tutorshop

**Identify regularities and challenges (data mining studies, secondary analysis)**
- KCs learning in equation solving [36]
- Effect of teacher interventions [25, 26]
- investigations of variability in learning rate [30]

**Identify opportunities for improvement, redesign**
- Multi-method approach [24]

**Intervention studies**
- **Within-tutor studies (i.e., test learning principles in the context of anITSs)**
  - Graphical representations, self-explanation prompts [54, 62]
  - Worked examples [17; 41, 59])
  - Strategy freedom [60]
  - Gamification [33, 44]
  - Polite language [39]
  - Blocking v. interleaving [48, 51]
  - Sequencing of sense-making v. fluency-building activities [50, 53]
  - Grounded feedback [61]
  - Higher v. lower scaffolding [5, 8]
  - High-similarity v. reduced-similarity example-problem pairs [9]
  - ITSs for students too young to read [56]
  - Dictionary skill development [37]
  - Diagrammatic self-explanation [43]
  - Self-regulated learning (self-assessment, problem selection, use of optional graphical representations) [34, 44]
  - Combining individual v. collaborative learning supported by ITSs [49, 66]
  - Guided invention activities [12, 55]
- **ITS v. non-ITS control condition**
  - Tutor v. educational game [35]
  - tutor v. problem solving on paper [7]
  - Adding tutoring to a simulation [58]
- **Teacher or student dashboards**

> - ○ Teacher support tools [19, 65]
> - ○ Student agency [18]
> - **Design-loop adaptivity**
>   - ○ Data-driven redesign based on KC model refinement [24]
>
> **Miscellaneous**
> - Effectiveness of methods for data-driven tutor improvement [24]
> - Effectiveness of authoring tools to build ITSs [1]

As of June, 2024, the total number of datasets in DataShop coming from CTAT tutors equals 1,082; the vast majority (97%) of these are from 2010 or later. This number, however, overestimates the number of studies done with CTAT tutors for several reasons: First, sometimes the data from different sites where a study is run (e.g., schools) are captured in separate DataShop datasets so that multiple datasets are associated with a single study. Second, sometimes a dataset is used solely for testing a tutor prior to use in a study. Third, in the analysis phase, datasets are often duplicated or partially duplicated, e.g., as part of KC modeling efforts. Fourth, sometimes a dataset is created for reasons other than research, for example, for prototyping or regular use of a tutoring system or for researchers exploring the tools.

We try to (imperfectly) correct for these possibilities by: (1) only counting datasets that have at least 300 transactions, on the assumption that datasets generated from tutor testing tend to have a small number of transactions; (2) only counting datasets that do not have the word "test" or the word "pilot" in their name; and (3) counting at most one dataset per semester per DataShop project, on the assumption that studies rarely cross semester boundaries and that multiple concurrently-collected datasets in a project are likely from a single research effort. (A "project" is DataShop's top-level organization, associating one or more datasets to a group of researchers.) With these restrictions, the number of datasets from CTAT tutors is reduced to 294. This number includes both experimental studies and data from regular use of the tutor. To adjust for other uses of the platform, we estimate that the number of experimental studies is likely no more than half this number (147).

To estimate the number of researchers who have been active on the platform, we count the number of accounts authorized to create tutors on Tutorshop and, of those accounts, the ones that actually did so (i.e., created new tutors, tutor variations, or problem sets) (see Table 2). These accounts include those with explicit Experimenter privileges but also include Developer accounts (researchers typically had one or the other). These numbers might overestimate the number of active researchers on the platform, because they include accounts that were involved in creating tutors for regular instructional use. Since we cannot easily separate these two categories, we will therefore (again, imperfectly) correct for the overestimation by conservatively assuming that only half the accounts that had content creation privileges were used by researchers. As a criterion for creating content, we check if the account created at least one "package" on Tutorshop; the files for tutoring systems on Tutorshop are organized into packages, and new tutors, tutor features, or problem sets are likely to be found in separate packages.

The number of accounts that created content on CTAT+Tutorshop is greater than 800 over the lifetime of the platform (see Table 2). We estimated this number by counting the number of Tutorshop account holders who (known to us) conducted research studies on Tutorshop, created tutors for regular use in courses, attended tutor-building courses or workshops, or who contacted us via the "Contact Us" page and then went on to use the system, sometimes with very little involvement by us. If we restrict the number to the past year only, it is equal to 121. As discussed, the number of *researchers* active on the platform might have been half these numbers (i.e., ~60 in the last year, and ~400 over the lifetime of the platform). We note, further, that not all accounts that had the privilege to create content actually did so. Finally, over the lifespan of the system, the total number of accounts created on the system is greater than 40k. The great majority of these accounts (95%) are student accounts, while over 4k accounts have authoring privileges (11%). Given that CTAT tutors can be deployed on platforms other than Tutorshop, the number of student accounts on Tutorshop undercount the real number. In particular, CTAT tutors used on the OLI platform are used by large numbers of students, many hundreds per semester.

**Table 2:** Adoption of CTAT+Tutorshop as expressed in user account numbers

|  | Since June 1, 2023 | Lifetime (*estimated) |
|---|---|---|
| Accounts that created at least 1 package | 121 | *800+ |
| Accounts with the privilege to create content (regardless of whether they actually created content) | 272 | 1,440 |
| Total accounts created, excluding student accounts | 401 | 4,726 |
| Total accounts created, including student accounts | 1,817 | 41,550 |

# 5   Case Studies

We present five cases of research studies carried out with CTAT+Tutorshop; between them they illustrate a range of research questions and research designs. They also illustrate the many ways that CTAT+Tutorshop are helpful in conducting research.

**Case study 1: Use of CTAT+Ttuorshop to test learning sciences principles in the context of an ITS**
A series of studies with a chemistry tutor, the *Stoich Tutor* [39-41], illustrates how CTAT+Tutorshop can be used without extensions to conduct research studies with ITSs. The *StoichTutor,* built with CTAT, provides learning support for high school students learning the sub-area of chemistry called stoichiometry. Key motivations were to provide better learning support in introductory chemistry courses but also to serve as a platform for learning sciences research. This tutor was used to explore several e-Learning principles [13], including the use of worked and erroneous examples for learning [41] and the impact of polite versus direct feedback for learning [39-40]. The conditions in these different experiments were implemented using standard features of the CTAT authoring tools. To create a polite language version, the researchers edited all hints, error messages, and success messages. Worked examples were created by having the tutor fill out some of the steps of standard problems, through a CTAT feature called tutor-performed actions, by which the tutor (rather than the student) updates the interface. What was left for the students to do was explain the terms in the solution equation using simple menus (an interaction already in the tutor). Finally, erroneous examples were built as their own kind of tutor problems, where a problem and erroneous solution were presented and the student was asked to explain and correct the error. The *Stoich Tutor,* like many tutors described in this paper, was deployed in TutorShop with DataShop as the data repository. The tutor variants in these studies illustrate why having authoring tools within a platform for learning sciences research is useful. The *Stoich Tutor* was used by later researchers [5] to address other research questions, illustrating another advantages of authoring tools in a platform for learning sciences research: tutors accumulate, so researchers have a choice.

**Case study 2: CTAT-built ITS v. Non-ITS control condition**
A study by [7] illustrates that CTAT+TutorShop can facilitate classroom experiments that compare learning with an ITS against a non-ITS control condition, in this case, problem-solving practice on paper without technology support. The case study also illustrates that methods often used to analyze log data from ed tech can sometimes be fruitfully applied to data from non-tech learning environments. The motivation for the study was to investigate the learning benefits of immediate feedback and as-needed instruction through hints (as is typical of an ITS) against a naturalistic homework control condition. The study employed a matched, within-subject design. It leveraged an existing CTAT-authored ITS for middle-school mathematics, Mathtutor (built on Tutorshop); specifically, the study focused on three units that deal with the topic of graph interpretation. Working with some of the teachers who participated in the study, the researchers created analogous paper versions of the tutor problems to serve as the control condition. The experimental conditions were set up in Mathtutor by having students work on two sets of Mathtutor units in a crossover fashion. The control condition (paper) was administered separately (i.e., outside of Mathtutor). Students were randomly assigned to conditions at the class level. The sequence of content units was fixed and was the same in each condition. Students started with either paper- or tutor-based practice and switched to the other condition for the second set of content. The Mathtutor/Tutorhop LMS facilities were used to assign classes their relevant content units each day. CTAT's capabilities for authoring tests were used to create pre/post test problems. A regular spreadsheet was used to randomly assign classes to conditions. An interesting challenge was converting the paper data from the control condition to DataShop format so the learning processes could be compared between the two conditions. Two human coders entered students' problem-solving steps on paper into a slightly modified version of the given tutor units so as to generate log data. This methodology allows for novel modeling of learning rates during problem solving on paper as a key way of studying the relative affordances of ITSs versus paper.

**Case study 3: Secondary analysis of log data from CTAT tutor**
Our next case study illustrates how CTAT and DataShop data can be used for secondary data analyses, specifically, to investigate the variability in students' learning rate. The researchers, Koedinger and collaborators [30] set out to identify fast learners in educational settings, for which they used 27 DataShop datasets from K-12 and college courses using AI Tutors, Online Courses, and Ed Games. Over half of these datasets were from CTAT tutors. A key strength of these datasets, and indeed of many of the datasets in DataShop, is the availability of cognitive models (KC models). Koedinger et al. used them to model student learning, applying a variation of the Additive Factors Model [11], a logistic growth model widely used in the field of Educational Data Mining that predicts the increase in accuracy resulting from each additional opportunity to practice a KC,. These analyses provided evidence to support the hypothesis that the path to expertise requires extensive practice: Students needed about 7 additional opportunities per KC in addition to lectures to master each KC. However, these analyses also provided evidence that, contrary to previous theoretical proposals, learners acquire competence at largely similar rates. The authors observed high variability in how easy or hard it is to learn each KC and noted that learners varied substantially in prior knowledge, but not learning rate. The results of this secondary data analysis investigation, using the detailed data and knowledge components available in DataShop made possible by CTAT and online courses, suggests that achievement gaps may largely be the result of opportunity gaps (different prior knowledge) but not learning rate; in other words, provided with enough opportunities in a high-support learning environment (such as an ITS) every learner can learn anything. However, the results also pose challenges to learning theory to explain the regularity in learning rate: Why are students so similar in learning rate? One proposal, still untested, is that different students might be pulling from different backgrounds, domain-general knowledge to support their learning leading to similar learning rates despite varied initial knowledge.

**Case study 4: Educational game v. simple tutor**
CTAT, Tutorshop, and DataShop have also been used to support the development of and experimentation with game-based learning environments. Since 2014, McLaren and colleagues have experimented with a learning game, *Decimal Point*, focused on middle school children learning decimals and decimal operations. Over the past decade, McLaren's lab has run a series of classroom experiments involving over 1,500 students with the game. The studies have included a comparison of the game to a standard tutoring system, a study of whether providing agency can lead to more learning and enjoyment, and whether students benefit from hints and feedback in the game. The most fundamental findings from this line of research are that *Decimal Point* has led both to more learning and enjoyment than the tutor and the game has led to a gender effect in which female students learn more from the game than male students. The game was developed using CTAT and involved writing a small amount of custom code to create mini-game templates of combined CTAT widgets to implement the five general problem types the game targets: addition of decimals, number lines, sorting, completing sequences, and putting decimals into less-than and greater-than buckets. The code written for the mini-games involved game mechanics and animations that were used to create a game-like context. The game is deployed on TutorShop and data is collected and analyzed in the DataShop. TutorShop has enabled ease of deployment at schools, while the DataShop has supported a fine-grained analysis of student learning beyond test scores [46] and a knowledge component analysis that has resulted in a deeper understanding of students' decimal skills. Similar effects were observed in Nagashima et al. [44], who compared a playful, gamified CTAT-based algebra tutor vs. a standard version of the tutor and found that the gamified tutor achieved greater student learning and engagement. Similarly, Long and Aleven [33] experimented with gamification features, implemented in a CTAT tutor (Lynnette) through custom programming in Tutorshop, inspired by standard video game design patterns, such as a star system to indicate how well a problem was solved and the possibility to redo a problem to gain more stars. This combination turned out to be detrimental for learning, as a caveat on importing game design patterns into learning software and ITSs.

**Case study 5: Test of design-loop adaptivity approach**
Our final case study illustrates how CTAT+Tutorshop can support and test "design-loop adaptivity" (Aleven et al., 2017) through the full research cycle described above. CTAT+Tutorshop have in fact facilitated several new learning engineering workflows that realize design-loop adaptivity, combined with classroom experiments that evaluate their effectiveness [24, 32]. Here, we report the design-loop adaptivity example of [24]. Data from a CTAT-authored ITS, which had been collected in DataShop, was analyzed, the tutor units were then updated based on insights of this analysis, and finally a close-the-loop study was conducted to test, in a between-subjects classroom experiment, whether the data-driven redesign enhanced student learning. The key idea is the use of log data to identify difficult skills and then to create focused tasks that target these difficult skills effectively.

The approach tested by Huang et al. [24] involves a three-stage workflow that combines existing and new learning analytics and instructional design methods. It focused on three units in Mathtutor [3] in which students learn to write or explain algebraic expressions that model real-world situations. The redesigned ITS differed from the original ITS in three aspects: a more fine-grained KC model, new focused tasks targeting difficult skills, and a new problem selection algorithm. The data-driven redesign was carried out using past data of Mathtutor collected from DataShop. CTAT+Tutorshop facilitated the process of authoring new focused tasks in three main ways. First, CTAT's authoring facilities were used to create new problem types. Second, CTAT was used to add customized feedback messages for selected errors identified in the log data. Third, CTAT libraries were used to parse and evaluate algebraic expressions. To create and integrate a new task selection algorithm, the researchers made use of TutorShop's API for custom outer loops, described above. The new algorithm leveraged the standard student model, which provides probabilities that the student has mastered each KC targeted in the tutor [14]. In addition, pre- and post-test problems were generated with CTAT's test mode, which disables feedback and hints. Finally, Tutorshop was used to configure the order in which students needed to go through the materials (e.g., students were required to finish the pre-test before moving on to practice in the first session).

The experiment was conducted with high school Algebra 1 classes for one month, during two 40-minute normal class periods per week. Students were randomly assigned to two conditions within each class period. This randomization was realized through importing a pre-generated spreadsheet to Tutorshop which specifies the mapping between student IDs and Assignments, where different conditions correspond to different Assignments. Tutorshop's dashboard was used to monitor students' progress in real time(with DataShop logging for later analysis). Huang et al. [24] found that the redesigned tutor led to significantly better learning gains, reduced over- and under-practice, yielded steeper learning curves, and improved difficulty progression. In Huang et al. [22, 23], the log data from this experiment was further used to analyze students' gaming-the-system behaviors, illustrating that one research cycle can sometimes feed into the next.

# 6 Discussion and Conclusion

ITSs support a learning scenario that is common and important in many task domains: deliberate practice, with complex practice problems, to develop cognitive skills. A distinguishing characteristic of ITSs is that they provide step-level guidance and individualized mastery learning; few ed tech products offer this combination. Given the strong evidence that ITSs help students learn very effectively [10, 38], further research involving ITSs would be highly valuable; it would strengthen the scientific understanding of learning with these systems and help develop a strong learning engineering practice for ITS. The CTAT+Tutorshop platform, together with DataShop support an iterative cycle of design and research with ITSs, enabling researchers to develop ITSs and conduct studies as well as to get insights into learner data. What is distinctive about CTAT+Tutorshop among research-enabling ed tech platforms is that it supports ITS research and provides both ITS authoring tools and several existing ITSs built with these tools; this way researchers can create experimental variations of existing tutors or build new tutoring systems from scratch.

Based on the number of datasets in DataShop that come from CTAT tutors, the number of studies run on the platform may be as high as 147. The number of relevant research accounts on Tutorshop (121 over the past year, an estimated 800+ over the CTAT+Tutorshop's lifespan) further indicates substantial use – we do not, however, see a reliable way of estimating the number of studies based on this number without considerable effort and consultation. Perhaps half the accounts were used by researchers who actively conducted research on the platform.

Generally, CTAT+Tutorshop have supported a large variety of studies addressing important, interesting research questions. These studies have moved forward the state-of-the-art in the learning sciences regarding learning in highly-scaffolded environments, as well as the learning engineering of ITSs. As indicated by Table 1 and illustrated by five case studies, the CTAT+Tutorshop platform makes it possible to address research questions covering a full research cycle (identify challenges using data from the ITS, improve the design of the ITS, test whether the improved ITS indeed enhances student learning). The studies used a wide range of research designs, including studies with multiple tutor versions, sometimes with factorial designs, experiments comparing tutor v. non-tutor conditions (with the non-tutor conditions running outside of the platform), and studies of design loop adaptivity. In some studies, the experimental manipulations occurred in the tutor's step loop [40], in others they changed the task loop [51], and in yet others, they were situated in both loops [24]. In some studies, the experimental tutor versions were authored straight with the tools [40, 51], in others, they involved custom programming [20, 33, 43, 44, 46, 66]. In all studies, Tutorshop's LMS capabilities were helpful in administering the experimental conditions.

Several additional features would help improve the platform. First, it would help to record, both on Tutorshop and in the DataShop log stream, meta-data of studies (e.g., contact information for the investigators, research questions, description of conditions, link to the pre-registration, link to the tutors in Tutorshop and to the code base, if there is a separate code base outside of Tutorshop - e.g., when custom programming was needed), so researchers could more easily get an up-to-date impression of what kinds of questions have and have not been investigated on the platform. This information, together with search capabilities, might help researchers who contemplate running a study on the platform to select research questions and/or find past projects to learn from. Second, it would help to enhance the support for experiments within CTAT+Tutorshop, for example, by explicitly representing studies and conditions, and adding functionality so the platform itself could, e.g., randomly assign of students or classes to conditions, counterbalance test forms assigned to students, and support the authoring of tutor behaviors that are contingent on the student's condition. Third, it might help as well to offer multi-armed bandit testing within the platform, in order to take advantage of advantageous conditions already while a study is still running, possibly within both the step loop and the task loop. Fourth, it would be useful to develop research templates that illustrate how to create an experiment on the platform -- e.g., proven ways of implementing a 3-condition between-subjects design, a 2x2 between-subjects design, a cross-over within-subjects design, etc. Although this wishlist is modest, further feedback from the research community could help generate additional ideas for how to strengthen CTAT+Tutorshop as a platform for research.

Regarding future research on the CTAT+Tutorshop platform, we see many open or under-researched questions for which the platform would be helpful. To mention a few, it would be interesting to study the value of the step loop, characteristic of ITSs, for learning outcomes. Very few such studies do so, a notable exception being a study by Corbett et al. [14] which showed that immediate feedback on the steps of problems supports both more effective and more efficient student learning. Given how foundational that result is with respect to both understanding learning in highly-scaffolded environments and engineering ITS designs, replication studies are highly desirable. To run this kind of study, one might create versions of the same ITS with and without a step loop (i.e., with the usual step-level guidance vs. giving feedback only at the end of the problem), or vary specific step loop features separately (prompts, feedback, and hints related to steps). Alternatively, one could separately design multi-step problems v. single-step problems for the same learning objectives. (The former would have a step loop, the latter would not.) It would also also be interesting to study the value (for students' learning outcomes) of the many possible forms of adaptivity within ITSs - as a way of addressing learning sciences questions about differentiation and personalization of instruction. The Adaptivity Grid [2] could help guide the brainstorming of research questions.

In closing, platform-enabled research is a great idea; the more platforms, the greater the variety in the research questions that can be studied. The CTAT+Tutorshop platform supports a full research lifecycle to answer learning sciences questions about learning in highly-scaffolded environments. Given that such research is valuable to the engineering of such environments, ultimately, it will mean that more students will more often use the best that ed tech has to offer.